\documentstyle[12pt]{article}

\begin{document}
\title{Invalid Ward-Takahashi identities and broken unitarity of the SM }
\author{Bing An Li\\Department of Physics, Univ. of Kentucky, Lexington, 
KY, 40506, USA}
\maketitle
\begin{abstract}
It is found that in the SM the Ward-Takahashi(WT) identities of the axial-vector currents 
and the charged vector currents of fermions are invalid after 
spontaneous symmetry breaking. The spin-0 components of Z and W fields are 
revealed from the invalidity of these WT identities. The masses of these 
spin-0 components are at $10^{14}$GeV. They are ghosts. Therefore, unitarity 
of the SM after spontaneous symmetry breaking is broken at $10^{14}$GeV.
\end{abstract}
\section{Introduction}
The SM[1] is successful in many aspects.
It is widely believed that the unitarity and renormalizability of the SM 
have been proved. A brief review of the history of the proof of the untarity and 
renormalizablity of the SM is presented.
\begin{enumerate}
\item Because of U(1) local gauge invariance and WT identities QED is unitary 
and
renormalizable. In QED there is vector coupling between fermion and the gauge field. 
\item Comparing with QED, the gauge fields in QCD are nonabelian.
Using nonabelian gauge invariance and WT identities, in 1971 't Hooft[2] has
proved that Yang-Mills field theory is unitary and renormalizable.
In QCD between quarks and 
Yang-Mills fields there are vector couplings.
Because of gauge invariance the WT identities including fermions can be 
naturally derived. 
Therefore, QCD is unitary and renormalizable.
\item In 1969 B.W. Lee and J.L.Gervais[3]
have proved the linear $\sigma$ model with a symmetry breaking term
c$\sigma$ is unitary and renormalizable. It shows that the WT identities 
is not affected by
the symmetry breaking term.
\item In 1971 't Hooft[4] studied
a large set of different models in which local gauge invariance(including $SU(2)$)
is broken spontaneously.
The fields in all these models studied are massive,
charged or neutral, spin one bosons, photons, and massive scalar particles.
Feynman rules and WT identities are found.
The author concludes that "A renormalizable and unitary theory results, with
photons, charged massive vector particles, and additional scalar particles."
At the end of the paper fermions
are introduced as
\[{\cal L}_N=-\bar{N}(m+\gamma_\mu D_\mu)N.\]
The fermions have the same mass and there is vector current only. 
\item In 1972 B.W.Lee[5] has proved that Abelian gauge field with spontaneous 
symmetry breaking is renormalizable and unitary.
\item In the four papers By B.W Lee and Jean Zinn-Justin[6] 
renormalizability and unitarity of the 
theory of
Yang-Mills fields
with spontaneous symmetry breaking and scalar fields have been proved in detail. 
It shows
that "spontaneous symmetry breaking doesn't affect the Ward-Takahashi identities". 
Therefore,
the theory is renormalizable and unitary. No fermions are included in the
studies of these papers. 
In their fourth paper[6](paper IV) the authors addresses the fermion 
problem
"Recently Bardeen[7]
has given a discussion of renormalizability
of gauge theories with fermions. He has shown that the problem associated with
anomalies in fermion loops can be isolated from the general problem of 
renormalizability
and gauge invariance, and if fermion loop anomalies are absent, or canceled among 
them
selves in lowest order, the presence of fermions does not hinder the WT identities 
from being valid."
\end{enumerate}
The SM consists of gauga bosons, scalars, and fermions. 
The structure related to fermions is far more complicated than both QED 
and QCD. Between fermions and gauge bosons there are both vector and
axial-vector couplings. Among vector couplings besides the neutral vector 
couplings 
there are charged vector couplings between fermions and W 
bosons. In both QED and QCD there are
neither axial-vector couplings nor charged vector couplings.  
These new currents have not been studied in
't Hooft's[4] and Lee and Zinn-Justin's papers[6]. Anomaly affects WT 
identities. The anomalies are canceled out at the lowest order in the 
SM. According to Bardeen[7], in the SM anomaly doesn't affect the WT identities
before and after spontaneous symmetry breaking.

In Ref.[8] we have presented a study on the effect of the vacuum polarization of 
Z and W bosons by fermions. The study[8] shows that after taking the vacuum 
polarization by massive fermions into account, both  Z and W fields have a spin-0
components whose masses are about $10^{14}$GeV. They are ghosts. Therefore, 
after spontaneous symmetry breaking 
unitarity
of the SM is broken at the level of $10^{14}$GeV. This result is in serious 
contradiction with the belief that the unitarity of the SM has been proved in Refs.
[4,6].  
In this paper we try to study the origin of the problem. 
The key point is that besides the anomaly
are there other effects of the two new kinds of fermion currents, mentioned above,
on the WT identities? 

The paper is organized as 1)introduction; 2) WT identities of fermions; 
3) vacuum polarization before spontaneous symmetry breaking; 4) vacuum 
polarization after spontaneous symmetry breaking; 5) conclusions.
\section{Ward-Takahashi identities of fermions}
Without losing generality, we start the study from t- and b- quark generation.
The Lagrangian of the SM for t and b quark generation is
\begin{eqnarray}
\lefteqn{{\cal L}=
-{1\over4}A^{i}_{\mu\nu}A^{i\mu\nu}-{1\over4}B_{\mu\nu}B^{\mu\nu}
+\bar{q}_L\{i\gamma\cdot\partial
+{g\over2}\tau_{i}
\gamma\cdot A^{i}+g'{Y\over2}\gamma\cdot B\}
q_{L}}\nonumber \\
&&+\bar{q}_{R}\{i\gamma\cdot\partial+g'{Y\over2}\gamma\cdot Bi\}q_{R}\nonumber \\
&&+f_t\{\bar{q}_L t_R\phi_c+\phi_c^\dag\bar{t}_R q_L\}
+f_b\{\bar{q}_L b_R\phi+\phi^\dag\bar{b}_R q_L\},
\end{eqnarray}
where
\[q_L=\left(\begin{array}{c}
             t_L\\b_L
            \end{array}  \right).\] 
The couplings between fermions,
Z, and W are
\begin{equation}
{\cal L}={\bar{g}\over4}\{
(1-{8\over3}\alpha)\bar{t}\gamma_\mu t+
\bar{t}\gamma_\mu\gamma_5  t\}Z^\mu
-{\bar{g}\over4}\{(1-{4\over3}\alpha)\bar{b}\gamma_\mu b
+\bar{b}\gamma_\mu\gamma_5 b\}Z^\mu,
\end{equation}
where \(\alpha=sin^2 \theta_W \),
\begin{equation}
{\cal L}={g\over2\sqrt{2}}\bar{t}\gamma_\mu(1+\gamma_5)b W^{+\mu}+
+\bar{b}\gamma_\mu(1+\gamma_5)t W^{-\mu}.
\end{equation}
Comparing with QED and QCD, the axial-vector currents and the charged vector currents 
are new.

Before spontaneous symmetry breaking the Lagrangian(1) is invariant under  
$SU(2)_L\times U(1)$ gauge transformation.
From Eq.(1) following equations of fermion currents are found
\begin{eqnarray}
\lefteqn{\partial_\mu(\bar{t}\gamma_\mu t)=\frac{i}{2\sqrt{2}}g\{\bar{t}\gamma_\mu(1+
\gamma_5)bW^-_\mu-\bar{b}\gamma_\mu(1+\gamma_5)tW^+_\mu\}},\nonumber \\
&&\partial_\mu(\bar{b}\gamma_\mu b)=\frac{i}{2\sqrt{2}}g\{\bar{b}\gamma_\mu(1+
\gamma_5)tW^+_\mu-\bar{t}\gamma_\mu(1+\gamma_5)bW^-_\mu\},\nonumber \\
&&\partial_\mu(\bar{t}\gamma_\mu t+\bar{b}\gamma_\mu b)=0.
\end{eqnarray}
The axial-vector currents satisfy following equations in unitary gauge
\begin{eqnarray}
\lefteqn{\partial_\mu(\bar{t}\gamma_\mu\gamma_5 t)=\frac{i}{2\sqrt{2}}g\{\bar{t}
\gamma_\mu\gamma_5 bW^-_\mu-\bar{b}\gamma_\mu\gamma_5 tW^+_\mu\}+i\sqrt{2}f_t\bar{t}
\gamma_5 t\phi^0+anomaly,}\nonumber\\
&&\partial_\mu(\bar(b)\gamma_\mu\gamma_5 b)=\frac{i}{2\sqrt{2}}g\{\bar{b}
\gamma_\mu\gamma_5 tW^-_\mu-\bar{t}\gamma_\mu\gamma_5 bW^+_\mu\}+i\sqrt{2}f_b\bar{b}
\gamma_5 b\phi^0+anomaly.
\end{eqnarray}
It is well known when adding up all quarks and leptons, the anomalies are 
canceled out in the SM. Because of this cancellation we will not mention anomalies again
in this paper.
The equations of charged vector and axial-vector currents are derived
\begin{eqnarray}
\lefteqn{\partial_\mu(\bar{t}\gamma_\mu(1+\gamma_5) b)=-ie\bar{t}\gamma\cdot Ab
-i\bar{g}(1-\alpha)\bar{t}\gamma_\mu(1+\gamma_5)bZ_\mu}\nonumber \\
&&+\frac{i}{\sqrt{2}}g\{\bar{t}\gamma_\mu(1+\gamma_5)t-\bar{b}\gamma_\mu(1+\gamma_5)
b\}W^+_\mu-{i\over\sqrt{2}}(f_b-f_t)\bar{t}b\phi^0
+{i\over\sqrt{2}}(f_b+f_t)\bar{t}\gamma_5 b\phi^0.
\end{eqnarray}

Eqs.(4-6) are the WT identities in the form of operators. We apply these equations to
study the vacuum polarizations of Z and W fields 
at the order of $O(\bar{g}^2)$ and 
$O(g^2)$ respectively. 

In Refs.[4,6] the unitarity of the theory of nonabelian gauge fields and scalar fields
with spontaneous symmetry breaking has been approved. In the SM fermion fields
are included
and there are two new kinds of currents: axial-vector and charged 
vector currents. According to Refs.[4.6] the vacuum polarizations of Z and W fields by
gauge bosons and Higgs don't have any effects on unitarity of the theory. Therefore,
we only study the vacuum polarizations by fermions in this paper. In the vacuum 
polarizations the two new currents are involved.
\section{Vacuum polarization before spontaneous symmetry breaking}
Before spontaneous symmetry breaking  
at the lowest order the equations(4-6) become
\begin{eqnarray}
\lefteqn{\partial_\mu\bar{t}\gamma_\mu t=0,\;\;\;\partial_\mu\bar{t}\gamma_\mu\gamma_5 t
=0,\;\;\;
\partial_\mu\bar{b}\gamma_\mu b=0,\;\;\;\partial_\mu\bar{b}\gamma_\mu\gamma_5 b
=0,}\nonumber \\
&&\partial_\mu(\bar{t}\gamma_\mu(1+\gamma_5)b)=0.
\end{eqnarray}
Before spontaneous symmetry breaking, the fermions of the SM are massless.
Using the vertices(2,3), the amplitudes of the vacuum polarization of Z and W fields
at one loop of fermions are calculated  
\begin{eqnarray}
\lefteqn{\Pi^Z_{\mu\nu}={\bar{g}^2\over8}\frac{N_C}{(4\pi)^2}D\Gamma(2-{D\over2})
\int^1_0 dx x(1-x)(\frac{\mu^2}{L_0})^{{\epsilon\over2}}(q_\mu q_\nu-q^2 g_{\mu\nu})
(y_t+y_b),}\nonumber\\
&&\Pi^W_{\mu\nu}={g^2\over4}\frac{N_C}{(4\pi)^2}D\Gamma(2-{D\over2})
\int^1_0 dx x(1-x)(\frac{\mu^2}{L_0})^{{\epsilon\over2}}(q_\mu q_\nu-q^2 g_{\mu\nu}),
\end{eqnarray}
where \(y_t=1+(1-{8\over3}\alpha)^2\), \(y_b=1+(1-{4\over3})^2\), 
\(L_0=m^2-x(1-x)q^2 \). 
m is a parameter introduced for the infrared divergence caused by massless 
quark. 
The amplitudes(8) show that the WT identities(7) are satisfied before spontaneous 
symmetry breaking.
The contributions of other generations of fermions can be obtained too. The WT 
identities are satisfied too.
\section{Vacuum polarization after spontaneous symmetry breaking}
After spontaneous symmetry breaking in unitary gauge there are
\begin{equation}
\phi^0=\eta+H,\;\;\;m_t=f_t \eta,\;\;\;m_b=f_b \eta.
\end{equation}
Substituting Eqs.(9) into Eqs.(4-6), new equations of fermion currents are obtained. 
After 
spontaneous symmetry breaking fermion mass appears 
in the equations of the axial-vector and charged vector currents.
The fermion mass originates in the violation of $SU(2)_L\times U(1)$ symmetry. 

The invalidity of the WT identities of fermion axial-vector and charged vector 
currents after 
spontaneous symmetry breaking can be shown clearly by the equations 
at the lowest order
\begin{eqnarray}
\lefteqn{\partial_\mu(\bar{t}\gamma_\mu\gamma_5 t)=2im_t\bar{t}\gamma_5 t,}\nonumber \\
&&\partial_\mu(\bar{b}\gamma_\mu\gamma_5 b)=2im_b\bar{b}\gamma_5 b,\nonumber \\
&&\partial_\mu(\bar{t}\gamma_\mu(1+\gamma_5)b)=i(m_t-m_b)\bar{b}t+i(m_t+m_b)
\bar{t}\gamma_5 b.
\end{eqnarray}

In the models studied in Refs.[4,6] there are no axial-vector and charged vector 
currents of fermions. Therefore, there are no such violations of WT identities(10). 

We need to study the physical effects caused by the invalid WT identities(10). 
In this paper 
the vacuum polarization of Z and W fields at one loop of fermions are 
studied.

The amplitude of the vacuum polarization of Z fields by massive fermions 
at the lowest order is expressed as[8]
\begin{equation}
\Pi^Z_{\mu\nu}={1\over2}F_{Z1}(z)(p_\mu p_\nu-p^2 g_{\mu\nu})+F_{Z2}(z)
p_\mu p_\nu+{1\over2}\Delta m^2_Z g_{\mu\nu},
\end{equation}
\begin{eqnarray}
\lefteqn{F_{Z1}=1+\frac{\bar{g}^2}{64\pi^2}\{\frac{D}{12}\Gamma(2-{D\over2})
[N_C y_q\sum_q(\frac{\mu^2}{m^2_q})^{{\epsilon\over2}}
+y_l\sum_l(\frac{\mu^2}{m^2_l})^{{\epsilon\over2}}]}\nonumber \\
&&-2[N_C y_q\sum_q f_1(z_q)+y_l\sum_lf_1(z_l)]+2[\sum_q f_2(z_q)
+\sum_{l=e,\mu,\tau}
f_2(z_l)]\},
\nonumber \\
&&F_{Z2}=-\frac{\bar{g}^2}{64\pi^2}\{N_C\sum_qf_2(z_q)
+\sum_{l=e,\mu,\tau}f_2(z_l)\},\\
&&\Delta m^2_Z={1\over8}{\bar{g}^2\over(4\pi)^2}D\Gamma(2-{D\over2})
\{N_c\sum_{q}m^2_q
({\mu^2\over m^2_q})^{{\epsilon\over2}}+\sum_l m^2_l({\mu^2\over m^2_l})^
{{\epsilon\over2}}\}\nonumber .
\end{eqnarray}
where \(y_q=1+(1-{8\over3}\alpha)^2\) for \(q=t,c,u\),
\(y_q=1+(1-{4\over3}\alpha)^2\)
for
\(q=b,s,d\), \(y_l=1+(1-4\alpha)^2\), for \(l=\tau, \mu, e\), \(y_l=2\) for
\(l=\nu_e,\nu_\mu,\tau_\mu\),
\(z_i={p^2\over m^2_i}\),
\begin{eqnarray}
\lefteqn{f_1(z)=\int^1_0dxx(1-x)log\{1-x(1-x)z\},}\nonumber \\
&&f_2(z)={1\over z}\int^1_0 dxlog\{1-x(1-x)z\}.
\end{eqnarray}

Comparing with Eq.(8), the expression of the vacuum polarization by massive fermions
has two new terms: $F_{Z2}(z)$ and $\Delta m^2_Z$. 
Both the vector and axial-vector couplings contribute to $F_{Z1}$.
Only the axial-vector currents of massive fermion(2) contribute to $F_{Z2}$ 
and $\Delta m^2_Z$. Eq.(12) show that in the limits, $m_{q(l)}\rightarrow 0$, 
we have
\[\Delta m^2_Z=0,\;\;\;F_{Z2}(z)=0.\]
Therefore, the two new terms are resulted in the violation of the WT identity of the 
axial-vector currents of fermions of the SM after spontaneous symmetry breaking.
Because of these two new terms there is no current conservation in Eq.(11).  
The function $F_{Z1}$ is used to renormalize the Z-field.
In the SM the Z boson gains mass from the spontaneous symmetry breaking and
$\Delta m^2_Z$ can be refereed to the renormalization of $m^2_Z$.
$F_{Z2}$ is finite and
it indicates that Z field has nonzero divergence. 
$\partial_\mu Z^\mu$ can be defined as a spin-0 field $\phi_Z$ whose mass
is labeled as $m_{\phi_Z}$ which can be determined. Therefore, 
the Z-field has the fourth independent 
spin-0 component $\phi_Z$. 
We rewrite $F_{Z2}$ as
\begin{equation}
F_{Z2}(z)=\xi_Z+(p^2-m^2_{\phi_Z})G_{Z2}(p^2),
\end{equation}
$G_{Z2}$ is the radiative correction, and
\begin{equation}
\xi_Z
=F_{Z2}|_{p^2=m^2_{\phi_Z}}.
\end{equation}
The new free Lagrangian of the Z-field
is constructed as
\begin{equation}
{\cal L}_{Z0}=-{1\over4}(\partial_\mu Z_\nu-\partial_\nu Z_\mu)^2
+\xi_Z(\partial_\mu Z^\mu)^2+{1\over2}m^2_Z Z^2_\mu.
\end{equation}
The term $(\partial_\mu Z^\mu)^2$ is dynamically generated and the coefficient $\xi_Z$
is no longer a parameter of gauge condition and it can be determined
(see the result below). It is necessary to emphasize that at $O(\bar{g}^2)$ the term
$\xi_Z(\partial_\mu Z^\mu)^2$ of Eq.(16) is gauge independent.
The new spin-0 field is defined as[8]
\begin{equation}
\phi_Z=\mp{m_Z\over m^2_{\phi_Z}}\partial_\mu Z^\mu.
\end{equation}
The equation of $\phi_Z$ is derived from Eq.(16) as
\begin{equation}
\partial^2\phi_Z-{m^2_Z\over2\xi_Z}\phi_Z=0.
\end{equation}
Therefore, the mass of $\phi_z$ is defined as
\begin{equation}
m^2_{\phi_Z}=-{m^2_Z\over2\xi_{Z}}.
\end{equation}
Using the Eq.(15), $m^2_{\phi_Z}$ satisfies
\begin{equation}
2m^2_{\phi_Z}F_{Z2}|_{p^2=m^2_{\phi_Z}}+m^2_Z=0.
\end{equation}
From the expression(12), the value of $m^2_{\phi_Z}$ is determined by
\begin{equation}
3\sum_q{m^2_q\over m^2_Z}z_q f_2(z_q)+\sum_l {m^2_l\over m^2_Z}z_l
f_2(z_l)={32\pi^2\over\bar{g}^2}.
\end{equation}
For $z>4$ 
\begin{equation}
f_2(z)=-{2\over z}-{1\over z}
(1-{4\over z})^{{1\over2}}log\frac{1-(1-{4\over z})^{{1\over2}}}
{1+(1-{4\over z})^{{1\over2}}}
\end{equation}
is a very good approximation.
Because of the ratios of ${m^2_q\over m^2_Z}$ and ${m^2_l\over m^2_Z}$
top quark dominates and the contributions of other fermions
can be
ignored. The equation has a solution at very large value of z. For very large z
we have
\begin{equation}
\frac{2(4\pi)^2}{\bar{g}^2}+\frac{6m^2_t}{m^2_Z}=3{m^2_t\over m^2_Z}log
{m^2_{\phi_Z}\over m^2_t}.
\end{equation}
The mass of the $\phi_Z$ is determined to be
\begin{equation}
m_{\phi_Z}=m_t e^{\frac{m^2_z}{m^2_t}{16\pi^2\over3\bar{g}^2}+1}
=m_t e^{28.4}=3.78\times10^{14}GeV,
\end{equation}
and
\[\xi_Z=-1.18\times10^{-25}.\]
The neutral spin-0 boson is extremely heavy.
The Z-filed is decomposed as
\begin{eqnarray}
\lefteqn{Z_\mu=Z'_\mu\pm{1\over m_Z}\partial_\mu\phi_Z,}\\
&&\partial_\mu Z'^\mu=0.
\end{eqnarray}

Using the Lagrangian(16),
the propagator of Z boson is found
\begin{equation}
\Delta_{\mu\nu}=
\frac{1}{p^2-m^2_Z}\{-g_{\mu\nu}+(1+\frac{1}{2\xi_Z})\frac{p_\mu p_\nu}{
p^2-m^2_{\phi_Z}}\},
\end{equation}
It can be separated into two parts
\begin{equation}
\Delta_{\mu\nu}=
\frac{1}{p^2-m^2_Z}\{-g_{\mu\nu}+\frac{p_\mu p_\nu}{
m^2_Z}\}-\frac{1}{m^2_Z}\frac{p_\mu p_\nu}{p^2
-m^2_{\phi_Z}}.
\end{equation}
The first part is the propagator of the physical spin-1 Z boson and
the second part is the propagator of a new
neutral spin-0 meson, $\phi_Z$.

The minus sign of Eq.(28) indicates that
the Fock space has indefinite metric and there is problem of
negative probability when $\phi_Z$ is on mass shell. 
The $\phi_Z$ is a ghost field. Therefore, unitarity of the SM
is broken at \(E=m_{\phi_Z}\). This ghost field is dynamically generated by fermion loops
and 
there is no way it can be canceled.

Similarly, the expression of the vacuum polarization of W-fields at one loop of fermions 
is calculated[8]
\begin{equation}
\Pi^W_{\mu\nu}=F_{W1}(p^2)(p_\mu p_\nu-p^2 g_{\mu\nu})+2F_{W2}(p^2)
p_\mu p_\nu+\Delta m^2_W
g_{\mu\nu},
\end{equation}
where
\begin{eqnarray}
\lefteqn{F_{W1}(p^2)=1+{g^2\over32\pi^2}D\Gamma(2-{D\over2})\int^1_0 dx
x(1-x)\{
N_C\sum_{iq}({\mu^2\over L^i_q})^{{\epsilon\over2}}+
\sum_{il}({\mu^2\over L^i_l})^{{\epsilon\over2}}\}}\nonumber \\
&&-{g^2\over16\pi^2}\{N_C\sum_{iq} f^i_{1q}+\sum_{il}
f^i_{1l}\}+{g^2\over16\pi^2}\{N_C\sum_{iq} f^i_{2q}+\sum_{il}f_{2l}\},\\
&&F_{W2}(p^2)=-{g^2\over32\pi^2}\{N_C\sum_{iq} f^i_{2q}+\sum_{il}
f^i_{2l}\},\\
&&
\Delta m^2_W={g^2\over4}{1\over(4\pi)^2}D\Gamma(2-{D\over2})
\int^1_0 dx\{N_c\sum_{iq}L^i_q
({\mu^2\over L^i_q})^{{\epsilon\over2}}+\sum_{il}L^i_l({\mu^2\over L^i_l})^
{{\epsilon\over2}}\}.
\end{eqnarray}
where
\begin{equation}
L^1_q =m^2_b x+m^2_t (1-x),\;\;
L^2_q =m^2_s x+m^2_c (1-x),\;\;
L^3_q =m^2_d x+m^2_u (1-x),
\end{equation}
\[L^1_l =m^2_e x,\;\;
L^2_l =m^2_\mu x,\;\;
L^3_l =m^2_\tau x,\]
\begin{eqnarray}
\lefteqn{
f^i_{1q}=\int^1_0 dx x(1-x)log[1-x(1-x){p^2\over L^i_q}]},\\
&&f^i_{1l}=\int^1_0 dx x(1-x)log[1-x(1-x){p^2\over L^i_l}],\\
&&f^i_{2q}={1\over p^2}\int^1_0 dx L^i_q log[1-x(1-x)
{p^2\over L^i_q}],\\
&&f^i_{2l}={1\over p^2}\int^1_0 dx L^i_l log[1-x(1-x){p^2\over L^i_l}].
\end{eqnarray}

Unlike Eq.(8), the amplitude(29) doesn't satisfy current conservation after spontaneous
symmetry breaking. 
There are two new terms in the amplitude(29), $F_{W2}$
and $\Delta m^2_W$, which cause the violation of current conservation. 
These two new terms violate the WT identities(7). 
As shown in Eq.(10), the invalidity of the WT identity 
is 
caused by the charged currents of massive fermions. In fact, in the limit, 
$m_q(m_l)\rightarrow 0$, we obtain
\[F_{W2}=0,\;\;\;\Delta m^2_W=0.\]
The amplitude(29) goes back to the expression(8), current conservation is satisfied. 
Therefore, $F_{W2}$ and $\Delta m^2_W$ 
are
the results of the invalid WT identity(10).
The function $F_{W1}(p^2)$ is used to renormalize the W-field. $\Delta m^2_W$ can be 
refereed 
to the renormalization of $m^2_W$. $F_{W2}$ is finite and it makes nonzero divergence 
of W-fields. W fields have four independent components. According to Ref.[8], 
nonzero $\partial_\mu W^{\pm\mu}$ lead to the existence of two charged spin-0 states,
$\phi^\pm_W$, whose mass is labeled as $m_{\phi_W}$ which can be determined.
$F_{W2}$ is rewritten as[8]
\begin{eqnarray}
\lefteqn{F_{W2}=\xi_W+(p^2-m^2_{\phi_W})G_{W2}(p^2),}\\
&&\xi_W=F_{W2}(p^2)|_{p^2=m^2_{\phi_W}},
\end{eqnarray}
where $G_{W2}$ is the radiative correction of the term
$(\partial_\mu W^\mu)^2$. 

The free part of the Lagrangian of W-fields
is redefined as
\begin{equation}
{\cal L}_{W0}=-{1\over2}(\partial_\mu W^+\nu-\partial_\nu W^+_\mu)
(\partial_\mu W^-_\nu-\partial_\nu W^-_\mu)
+2\xi_W\partial_\mu W^{+^\mu}
\partial_\nu W^{-^\nu}\]
\[+m^2_W W^+_\mu W^{-\mu}.
\end{equation}
In Ref.[8] the new $\phi^{\pm}_W$ fields are found to be
\begin{equation}
\phi^\pm_W=\mp{m_W\over m^2_{\phi_W}}\partial_\mu W^{\pm\mu}.
\end{equation}
They satisfy the equation
\begin{equation}
\partial^2\phi^\pm_W-{m^2_W\over2\xi_{W}}\phi^\pm_W=0.
\end{equation}
Therefore, the mass of $\phi^{\pm}_W$ is defined as
\begin{equation}
m^2_{\phi^\pm_W}=-{m^2_W\over 2\xi_{W}}
\end{equation}
Eq.(39) becomes
\begin{equation}
2m^2_{\phi_W}F_{W2}(p^2)|_{p^2=m^2_{\phi_W}}+m^2_W=0
\end{equation}
Top quark dominates the Eq.(44)
\begin{equation}
{p^2\over m^2_W}F_{W2}=-\frac{3g^2}{32\pi^2}{m^2_t\over m^2_W}\{-{3\over4}
+{1\over2z}
+[{1\over2}-{1\over z}+{1\over2z^2}]log(z-1)\}.
\end{equation}
Eq.(45) has a solution at very large z
\begin{equation}
m_{\phi_W}=m_t e^{{16\pi^2\over3g^2}{m^2_W\over m^2_t}}=m_t e^{27}
=9.31\times10^{13}GeV
\end{equation}
and
\begin{equation}
\xi_W=-3.73\times10^{-25}.
\end{equation}

The propagator of W-field is derived from the Lagrangian(40)
\begin{equation}
\Delta^W_{\mu\nu}=
\frac{1}{p^2-m^2_W}\{-g_{\mu\nu}+(1+\frac{1}{2\xi_W})\frac{p_\mu p_\nu}{
p^2-m^2_{\phi_W}}\},
\end{equation}
and it can be separated into two parts
\begin{equation}
\Delta^W_{\mu\nu}=
\frac{1}{p^2-m^2_W}\{-g_{\mu\nu}+\frac{p_\mu p_\nu}{
m^2_W}\}-\frac{1}{m^2_W}\frac{p_\mu p_\nu}{p^2
-m^2_{\phi_W}}.
\end{equation}
The first part is the propagator of physical
spin-1 W-field and the second
part is the propagator of the $\phi^\pm_W$ fields.
The minus sign in front of the propagator of $\phi^\pm_W$ fields 
indicates that there is problem of indefinite metric and negative
probability when $\phi^{\pm}_W$ are on mass shell. Therefore, $\phi^\pm_W$ 
are ghost fields.
Unitarity of the SM is broken
at \(E\sim m_{\phi_W}\).

The study shows that the WT identities of axial-vector and charged vector currents 
are invalid
after fermions gain masses from the spontaneous symmetry breaking. 
The invalidity of these
WT identities after spontaneous symmetry breaking generate three spin-0 
ghosts whose masses
are at $10^{14}$ Gev. These ghosts are dynamically generated by fermion loops and they
cannot be canceled. Therefore, unitarity of the SM is broken at this energy level.

The author wish to thank S.Drell, S.Brodsky, H.Quinn, and M.Weinstein for discussion. 
This research
is supported by a DOE grant.

\end{document}